\documentclass[prl,aps,twocolumn,showpacs]{revtex4}
\begin{document}
\input{epsf}
\title{Symmetry-breaking instability
in a prototypical driven granular gas}
\author{Evgeniy Khain and Baruch Meerson }
\affiliation{Racah Institute  of  Physics, Hebrew University of  Jerusalem, Jerusalem 91904,
Israel}
\begin{abstract}
Symmetry-breaking instability of a laterally uniform granular cluster (strip state)  in a
prototypical driven granular gas is investigated. The system consists of smooth hard disks in a
two-dimensional box, colliding inelastically with each other and driven, at zero gravity, by a
``thermal" wall. The limit of nearly elastic particle collisions is considered, and granular
hydrodynamics with the Jenkins-Richman constitutive relations is employed. The hydrodynamic
problem is completely described by two scaled parameters and the aspect ratio of the box. Marginal
stability analysis predicts a spontaneous symmetry breaking instability of the strip state,
similar to that predicted recently for a different set of constitutive relations.  If the system
is big enough, the marginal stability curve becomes independent of the details of the boundary
condition at the driving wall. In this regime, the density perturbation is exponentially localized
at the elastic wall opposite to the thermal wall. The short- and long-wavelength asymptotics of
the marginal stability curves are obtained analytically in the dilute limit. The physics of the
symmetry-breaking instability is discussed.
\end{abstract}
 \maketitle
\section{I. Introduction}

Granular materials play an important role in industrial applications, geophysics and
astrophysics. They are also of a great general interest to physicists, as each of the ``phases" of
granular materials: solid, liquid and gas have unusual properties that distinguish them from their
classic (atomic) counterparts \cite{Jaeger,Duran}. We will consider granular gas (or rapid
granular flow) and focus on a variant of clustering instability, a striking tendency of granular
gases to form dense clusters. Clustering instability was first discussed in the context of a
freely "cooling" granular gas \cite{Hopkins,Goldhirsch,McNamara}. Following these works, related
clustering phenomena were investigated in driven granular gases as well, both in experiments
\cite{Kudrolli,Urbach} and in particle simulations \cite{Kadanoff2,Grossman,Esipov}.

Granular clustering results from energy losses by inelastic collisions, and it is a manifestation
of thermal condensation instability, also encountered in other fields, for example in  gases and
plasmas that cool by their own radiation \cite{plasma}. Since the discovery of the clustering
instability, the validity of granular hydrodynamics (see Ref. \cite{Campbell} for a review) has
been under scrutiny \cite{Kadanoff1}. In contrast to the clustering in a freely "cooling" granular
gas, where one deals with a complex time-dependent process, steady states are achievable in driven
granular systems. One of the simplest settings of this type is driving the granulate by a side
wall at a zero gravity. Therefore, an ensemble of inelastically colliding hard spheres, confined in
a box and driven by one or two ``thermal" walls has served as a prototypical driven granular system
\cite{Kudrolli,Kadanoff2,Grossman,Brey,Tobochnik,Kudrolli2,LMS}. Steady states of this system have
served as test beds for granular hydrodynamics and its violations. The first analysis of this
system in the physical literature was performed, in one dimension, by Kadanoff \textit{et al.}
\cite{Kadanoff2}. The nearly elastic particles were constrained to move on a straight line with
energy input from the boundaries. Particle simulations \cite{Kadanoff2} showed that, for typical
initial conditions, the system evolves to a state where the particles are separated into two
groups. Almost all particles form a cluster in a small region of space, where they move with very
small velocities, while a very few remaining particles move with high velocities. Clearly, this
steady state cannot be described by granular hydrodynamics (actually, by any coarse-grained
theory). Therefore, the results of Kadanoff \textit{et al.} \cite{Kadanoff2} brought into question
the validity of granular hydrodynamics in general.

This question was addressed in two subsequent theoretical works \cite{Grossman,Esipov} that dealt
with similar systems in two dimensions. Esipov and P\"{o}schel \cite{Esipov} investigated an
ensemble of nearly elastic hard disks in a circular box with the circumference serving as thermal
wall. Grossman \textit{et al.} \cite{Grossman} considered a rectangular box, one side of which
served as thermal wall. Particle simulations \cite{Grossman,Esipov} showed granular clusters: dense
and ``cold" regions of granulate developing away from the thermal wall. In terms of
thecoarse-grained particle density, these steady-state clusters had simple shapes: azimuthally
uniform (circular state) \cite{Esipov} and laterally uniform (strip  state) \cite{Grossman}.
Grossman \textit{et al.} also showed that, for nearly elastic collisions, the strip state is
describable by a steady-state solution of granular hydrodynamic equations. The empiric
constitutive relations suggested by Grossman \textit{et al.} used simple interpolations between
the low-density limit, where the constitutive relations are derivable systematically
\cite{dilute}, and high-density limit where, close to the dense packing, free volume arguments can
be used.

The results of Refs. \cite{Grossman,Esipov} showed that the anomaly observed in the one-dimensional
setting \cite{Kadanoff2} does not persist in higher dimensions. Clustered states qualitatively
similar to those of Grossman \textit{et al.} were observed in experiment of Kudrolli \textit{et
al.} \cite{Kudrolli} who investigated a system of spherical particles in a box, rolling on a
smooth surface and driven by a rapidly vibrating side wall. The number of particles served as the
control parameter in Ref. \cite{Kudrolli}. A dense cluster of the strip type was observed when the
number of particles was big enough, in much the same way as in Ref. \cite{Grossman}. The basic
physics of the strip state is simple and can be explained by the following hydrodynamic argument.
Because of the inelastic collisions the particle random motion slows down (that is, the granular
temperature decreases) with the increase of the distance from the driving wall. To maintain the
momentum balance, the granular density should increase with this distance. When the total number of
particles is big enough (the rest of parameters being the same), the density contrast becomes
large, and the enhanced density region away from the driving wall is observed as the strip state.

The prototypical system exhibits many interesting phenomena of \textit{non}-hydrodynamic nature.
These include inelastic collapse \cite{Esipov,Tobochnik}, possible lack of scale separation
\cite{Grossman}, non-Gaussianity in the particle velocity distribution \cite{Grossman,Kudrolli2},
normal stress difference and pressure nonuniformities \cite{Brey}, etc. For nearly elastic
collisions, however, granular hydrodynamics was shown to yield an accurate quantitative
description in the dilute limit \cite{Brey}, and reasonably accurate description for moderate and
high granular densities \cite{Grossman}. Of course, the nearly elastic limit is quite restrictive
for most of realistic granular flows. Still, this limit is conceptually important just because
granular hydrodynamics can be used there. Granular hydrodynamics has a great predictive power and
helps to identify important collective phenomena (shear flows and vortices, shocks, different
modes of clustering flows etc.) that are difficult, if not impossible, to identify and predict in
the language of individual particles. Once identified, these phenomena can then be investigated in
experiment and simulations in more general (not necessarily hydrodynamic) formulations.

Therefore, granular hydrodynamics provide a leading-order approach to a big ensemble of
nearly-elastically colliding grains.  This approach has been taken recently by Livne \textit{et
al.} \cite {LMS} who employed granular hydrodynamics for a stability analysis of the strip state
described above. The analysis revealed a spontaneous symmetry-breaking instability of the strip
state with respect to perturbations along the strip. Well within the instability region, the
clustered states with broken symmetry, found by a numerical solution of the steady state
hydrodynamic equation, are strongly localized in the lateral direction: most of the particles are
located in dense ``islands" \cite{LMS}. These results indicate that the prototypical system can
show a non-trivial behavior even in the leading-order, hydrodynamic limit. Indeed, this systems
can be put into the list of pattern-forming systems far from equilibrium \cite{Cross}.

The present work focuses on a more detailed stability analysis of this system. Our first objective
is to check to what extent the symmetry-breaking instability predicted in Ref. \cite{LMS} is
sensitive to the constitutive relations. Livne et al. \cite{LMS} employed the empiric relations
suggested by Grossman et al. \cite{Grossman}. Here we shall use the better known Jenkins-Richman
(JR) relations \cite{Jenkins}. While the relations of Grossman et al. are more accurate for high
densities (even including those close to the dense packing limit), the JR relations should work
better at low and intermediate densities. We shall see, however, that the marginal stability
curves, obtained with these two sets of relations, are not much different from each other. This
implies that the symmetry-breaking instability is robust. Our second objective is to get more
insight into the marginal stability problem and, where possible, to obtain analytic results. We
shall show that the marginal problem is equivalent to an eigenvalue problem of quantum mechanics.
An important finding here is a universal behavior of the marginal stability curves in the limit
when the density perturbations are strongly localized at the elastic wall opposite to the thermal
wall. In the dilute limit, we obtain analytically the short- and long-wavelength asymptotics of the
marginal stability curves and density eigenfunctions. We also give the physical interpretation to
the symmetry-breaking instability and to the density borders of the instability region.

The rest of the paper is organized as follows. In Sec. 2 we formulate the model and briefly
describe the strip state: the simplest steady state of the prototypical system. Section 3 presents
marginal stability analysis of the strip state and compares the results obtained for two different
sets of constitutive relations. More results on marginal stability, including some analytic
results in the dilute limit, are presented in Sec. 4. Section 5 includes a discussion and summary.

\section{II. Prototypical system and strip state}.

The prototypical driven granular system in two dimensions include inelastically colliding hard
disks of diameter $d$ and mass $m=1$, moving in a box with dimensions $L_x \times L_y$. The
gravity force is zero. Collisions of disks with the walls $x=0, y=0$ and $y=L_y$ are assumed
elastic. The wall $x=L_x$ is "thermal" wall: upon collision a particle is assigned a random
velocity taken from a Gaussian ensemble with temperature $T_0$. Energy input at the thermal wall
balances the energy dissipation due to inter-particle collisions, so the system can reach a steady
state. We shall parameterize the inelasticity of the particle collisions by a constant normal
restitution coefficient $r$ and work in the nearly elastic limit: $1-r^2 \ll 1$. In this limit,
the Navier-Stokes granular hydrodynamics is expected to be sufficiently accurate in a system with
a big number of particles and small Knudsen number. The possible steady states of the system are
described by the steady state versions of the momentum and energy balance equations:
\begin{equation}
p=const\;\;\; \mbox{and} \;\;\;\nabla\cdot(\kappa\nabla T)=I\,, \label{E1}
\end{equation}
where $p$ is the granular pressure, $T$ is the granular temperature, $\kappa$ is the thermal
conductivity and $I$ is the rate of energy losses by collisions. We assume that the number density
$n$ is not too big: $n/n_c \leq 0.5$, where  $n_c = 2/(\sqrt{3} d^2)$ is the (hexagonal) dense
packing density. This assumption enables us to employ the constitutive relations derived by Jenkins
and Richman \cite{Jenkins}. For the steady state problem, the required constitutive relations
include the equation of state $p = p (n, T )$ and relations for $\kappa$ and $I$ in terms of $n$
and $T$.

Let us introduce scaled coordinates: ${\mathbf{r}}/L_x \to \mathbf{r}$. In the new coordinates the
box dimensions are $1 \times \Delta$, where $\Delta=L_y/L_x$ is the aspect ratio. Introducing the
normalized inverse density $z(x,y)=n_c/n(x,y)$, one can rewrite the energy balance equation in Eq.
(\ref{E1}) in terms of $z(x,y)$:
\begin{equation}
\nabla\cdot(F(z)\,\nabla z)=\eta \, Q(z), \label{E4}
\end{equation}
where $F(z)= A(z)\,B(z)$,
\begin{eqnarray}
A(z)&=&\frac{G\left[1+\frac{9\pi}{16}\left(1+\frac{2}{3G}\right)^2\right]}{z^{1/2}(1+2G)^{5/2}}\,,\nonumber
\\
B(z)&=&1+2G+\frac{\pi}{\sqrt{3}}\frac{z(z+\frac{\pi}{16\sqrt{3}})}{(z-\frac{\pi}{2\sqrt{3}})^3}\,,\nonumber
\\
Q(z)&=&\frac{6}{\pi}\frac{z^{1/2}G}{(1+2G)^{3/2}}\,,\nonumber
\\
G(z)&=&\frac{\pi}{2\sqrt{3}}\frac{z-\frac{7\pi}{32\sqrt{3}}}{(z-\frac{\pi}{2\sqrt{3}})^2}\,,
\label{E5}
\end{eqnarray}
and $\eta=(2\pi/3)(1-r)(L_x/d)^2$. Notice that, for an arbitrary small but finite inelasticity
$1-r$, the dimensionless parameter $\eta$ can be made arbitrary large, if the system size $L_x$ is
large enough.  The parameter $\eta$ differs from the parameter ${\cal L}$ used by Livne \textit{et
al.} \cite{LMS} only by a numerical factor of order unity. Of most interest are regimes where
$\eta \gg 1$, see below.

The boundary conditions for Eq. (\ref{E4}) are determined by the properties of the particle-wall
interactions. At the elastic walls $x = 0$, $y = 0$ and $y = \Delta$ we should prescribe a zero
normal component of the heat flux. In terms of the inverse density $z$ we have $\nabla_n z = 0$ at
these three walls. Here index $n$ denotes the gradient component normal to the wall. The constant
temperature at the ``thermal" wall $x=1$ yields the condition $\partial z(x=1,y)/\partial y=0$. To
make the formulation of the problem complete, one more condition is needed. In experiment or
particle simulations, the number of particles $N$ is fixed. This yields a normalization condition:
\begin{equation}
\frac{1}{\Delta}\int_0^1\int_0^\Delta \frac{dx dy}{z(x,y)}=f\,, \label{E6}
\end{equation}
where $f=\langle n \rangle / n_c$ is the area fraction of the grains and $\langle n \rangle =
N/(L_x L_y)$ is the average number density of the grains.

Equations (\ref{E4})-(\ref{E6}) and the boundary conditions make a complete set. One can see that
the governing parameters of this system are the scaled parameters $\eta$, $f$ and $\Delta$. If the
system is infinite in the $y$-direction, only two governing parameters: $\eta$ and $f$ remain.
Notice that the steady-state {\it density} distributions are independent of the wall temperature
$T_0$, in contrast to the similar problem with gravity, where the gravity acceleration, combined
with $T_0$ and the (finite) system size in the direction of gravity, would make an additional
governing parameter.

\begin{figure}[ht]
\vspace{-0.3 cm} \center{\epsfxsize=7.5 cm \epsffile{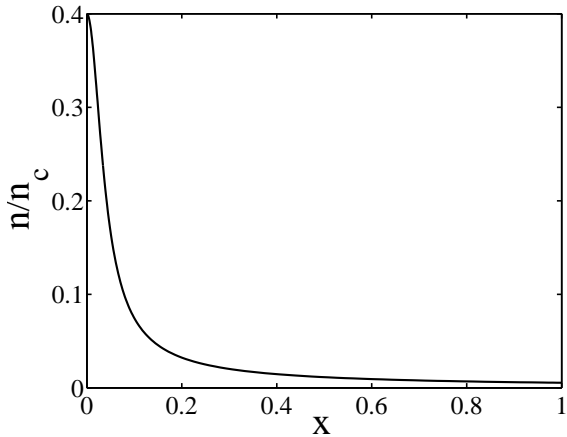}} \caption{An example of the scaled
density profile $n(x)/n_c$ of the strip state for the JR constitutive relations. The governing
parameters are $\eta=10^4$ and $f=0.0342$.} \label{fgr1}
\end{figure}

The laterally uniform steady state (strip state) corresponds to the one-dimensional
($y$-independent) solution $z = Z(x)$. It is described by the equations
\begin{eqnarray}
(FZ^{\prime})^{\prime}=\eta\, Q\,, \;\; Z^{\prime}|_{x=0}=0\,, \nonumber \\
\mbox{and} \;\;
\int_0^1 Z^{-1}(x)\,dx=f\,, \label{E7}
\end{eqnarray}
where primes stand for the $x$ derivatives. For the strip state, the boundary condition at the
wall $x=1$ drops out. This implies, in particular, that the \textit{density} profile of the strip
state is independent of the exact nature of the driving wall (thermal or vibrating wall)
\cite{boundary}. This degeneracy of the strip state is caused by the character of particle
interaction: the hard-core potential does not introduce any characteristic energy \cite{Esipov}.
Notice that, instead of prescribing the grain area fraction $f$, one can prescribe the inverse
density $Z=Z_0$ at $x=0$. This condition, combined with the no-flux condition at $x=0$ defines a
Cauchy problem for $Z(x)$. Solving the Cauchy problem, one can then compute, from the last
equation in Eq. (\ref{E7}), the respective value of $f$. At fixed $\eta$, there is a one-to-one
correspondence between $Z_0$ and $f$. Therefore, an alternative parameterization of the strip
state is given by the scaled numbers $\eta$ and $Z_0$. We shall see below that the same property
keeps (and can be conveniently used) in the marginal stability problem. Figure~\ref{fgr1} shows a
typical example of the scaled density profile $n(x)/n_c$ of the strip state obtained by solving
Eqs. (\ref{E7}) numerically.

\section{III. Instability of the strip state: marginal stability curves}

In general, the strip state is only one of the possible solutions of Eq. (\ref{E4}). Because of its
nonlinearity, Eq. (\ref{E4}) may have additional solutions satisfying the same boundary
conditions. When exist, these additional solutions are truly two-dimensional: the translational
symmetry along $y$ is broken. An important class of these solutions bifurcate supercritically from
the strip state \cite{LMS}. Therefore, close to the bifurcation point, these solutions can be found
by linearizing Eq. (\ref{E4}) around the strip state. A similar analysis was performed in Ref.
\cite{LMS} for the constitutive relations of Grossman \textit{et al.} \cite{Grossman}. In the
framework of a \textit{time-dependent} hydrodynamic formulation, this analysis corresponds to
\textit{marginal stability} analysis of the strip state with respect to small perturbations along
the strip \cite{LMS}.

\begin{figure}[ht]
\vspace{-0.3 cm} \center{\epsfxsize=7.5 cm \epsffile{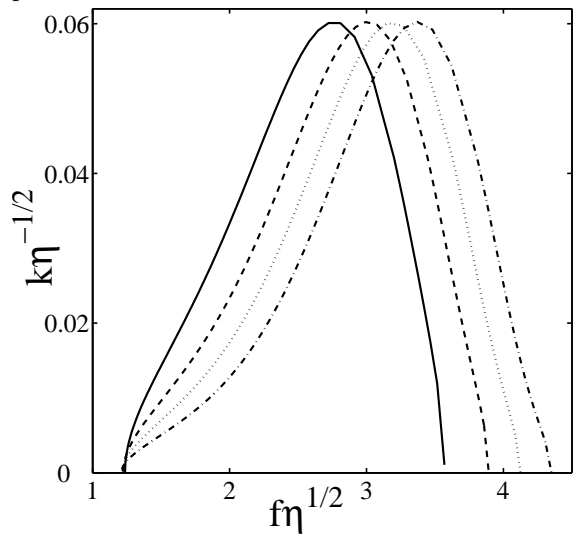}} \caption{Marginal stability curves
for different values of $\eta$, plotted in scaled coordinates: $k\eta^{-1/2}$ versus
$f\eta^{1/2}$.  For a fixed $\eta$ the strip state is stable above the respective curve and
unstable below the curve. The values of $\eta$ are: $10^4$ (solid line), $2.5\cdot 10^4$ (dashed
line), $5\cdot 10^4$ (dotted line) and $10^5$ (dash-dot line).} \label{f11}
\end{figure}

Substituting $z(x,y) = Z(x)+\psi_k(x)\cos ky$ and linearizing Eq. (\ref{E4}) with respect to the
small correction $\psi_k(x)\cos ky$, we obtain:
\begin{equation}
\phi^{\prime\prime}-\left(\frac{\eta Q_Z}{F}+k^2\right)\phi = 0\,. \label{E8}
\end{equation}
Here $\phi(x)=F(x) \,\psi_k(x)$, functions $F$ and $Q$ are evaluated at $z=Z(x)$, and subscript
$Z$ means the $z$ derivative evaluated at $z = Z(x)$. The boundary conditions are
\begin{equation}
\phi^{\prime}(x=0)=0\,\;\;\mbox{and}\;\;   \phi(x=1)=0\,. \label{E9}
\end{equation}
Equation (\ref{E8}) coincides with the Schr\"{o}dinger equation for an \textit{even} wave function
$\phi(x)$ of a particle in the potential well
\begin{equation}
U(x)=\left\{ \begin{array}{ll}
\frac{\eta\, Q_Z }{F}& \mbox{if}\;\; |x|<1\,,\\
+\infty& \mbox{otherwise}\,.
\end{array}
\right. \label{pot}
\end{equation}
The quantity $-k^2$ serves as the energy eigenvalue. The energy levels in the potential (\ref{pot})
are always discrete, and there is an infinite number of them. However, as the wave number $k$
should be real, only negative or zero energy levels are admissible. At fixed values of $\eta$ and
$f$, the potential (\ref{pot}) admits at most one such energy level. The absence of negative energy
levels implies that, in the vicinity of the strip state, there are no steady states different from
it. The presence of a negative energy level corresponds to a ``weakly two-dimensional" steady
state, bifurcating from the strip state. We shall exploit the quantum-mechanical analogy more
fully in Sec. 4. Here we report some numerical results. Figure~\ref{f11} shows the marginal
stability curves: the curves $k=k(f)$ at different values of $\eta$, computed numerically. In
these computations, the parameter $\eta$ was taken large enough. The strip state is unstable below
the respective curve and stable above the curve. Notice that, at fixed $\eta$, the instability is
possible only within a finite interval of $f$: $f_1(\eta) < f < f_2 (\eta)$. The same property was
reported in Ref. \cite{LMS} for another set of constitutive relations. We shall give a physical
explanation to this finding in Sec. 4. Notice (see also Ref. \cite{LMS}) that, at large $\eta$, the
high-density stability border $f_2$ is quite small. The curves in Fig.~\ref{f11} are actually
plotted in scaled coordinates: $k \eta^{-1/2}$ versus $f \eta^{1/2}$. It can be seen that, in the
scaled coordinates, all the curves exit from the same point of the horizontal axis $f \eta^{1/2}$.
In addition, the maxima of all the curves are equal. These observations will be also explained in
Sec. 4.

\begin{figure}[ht]
\vspace{-0.3 cm} \center{\epsfxsize=7.5 cm \epsffile{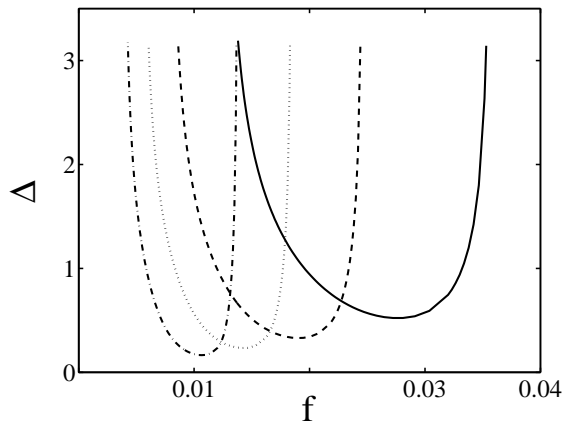}} \caption{The critical aspect ratio
$\Delta_1$ for the symmetry-breaking instability as a function of $f$ for different values of
$\eta$. For a fixed $\eta$, the strip state is stable below the respective curve and unstable
above the curve. The parameters are: $\eta_1=10^4$ (solid line), $\eta_2=2.5\cdot 10^4$ (dashed
line), $\eta_3=5\cdot 10^4$ (dotted line) and $\eta_4=10^5$ (dash-dot line). } \label{delta}
\end{figure}

If the system is infinite in the lateral direction, $\Delta = \infty$, while  $\eta$ and $f$ are
fixed, a whole continuum spectrum of wave numbers  between $k=0$ and $k=k (\eta,f)$ is unstable.
Both in experiment, and in numerical simulations $\Delta$ is finite. In this case $k$ becomes
discrete because of the boundary conditions: $k=m\pi/\Delta$, where $m=1,2,\dots$. For each $m$ we
can find the critical value of the aspect ratio $\Delta$ (let us call it $\Delta_m$) such that for
$\Delta>\Delta_m$ the strip state looses stability with respect to the $m$-mode. Obviously,
$\Delta_m = m \Delta_1$, and $\Delta_1$ is the lowest critical value for the symmetry-breaking
instability. Figure~\ref{delta} shows $\Delta_1$ as a function of $f$ for different values of
$\eta$. For fixed $\eta$ and $f$, the strip state is unstable above the curve. It is seen from
Fig. ~\ref{delta} that, in order to observe the symmetry-breaking instability, one does not need
to work with very large aspect ratios $\Delta$: it is sufficient if the system is big enough, so
that parameter $\eta$ is sufficiently large.

\begin{figure}[ht]
\vspace{-0.3 cm} \center{\epsfxsize=7.5 cm \epsffile{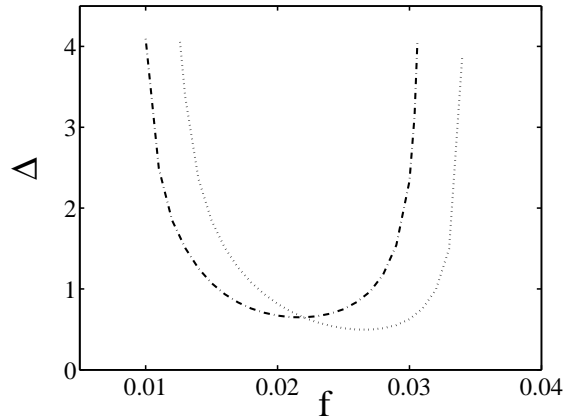}} \caption{The critical aspect ratio
$\Delta_1 (f)$ for the symmetry-breaking instability as computed for the constitutive relations of
Grossman et al. \protect \cite{Grossman} (dash-dotted line) and of JR \protect\cite{Jenkins}
(dotted line). Parameter $\eta=11,094$.} \label{delta1}
\end{figure}

To what extent is the symmetry-breaking instability sensitive to the precise form of the
constitutive relations? We compared the marginal stability curves $\Delta=\Delta_1(f)$ for
different values of $\eta$ with the respective curves \cite{LMS} found for the constitutive
relations of Grossman \textit{et al.} \cite{Grossman}. A typical example of this comparison is
shown in Fig.~\ref{delta1}. One can see that, qualitatively, the results are the same: the both
curves describe a symmetry-breaking instability at a critical value of the aspect ratio that
depends on the area fraction. In both cases, there are sharp low- and high-density borders of
instability region. Therefore, we can conclude that the instability is robust and does not require
a very special form of the constitutive relations. On the other hand, there is a noticeable (about
15\%) difference in the exact positions of the marginal stability curves, so the instability
provides a good quantitative test for constitutive relations of granular hydrodynamics.

\section{IV. More results on marginal stability}
In this Section we investigate the marginal stability problem in more detail and obtain some
analytic results in the dilute limit.
\subsection{A. Localization and universality}

Let us characterize the strip state by the scaled parameters $\eta$ and $Z_0$ and introduce a
different rescaling of the coordinate: $\bar{x}=x\,\eta^{1/2}$. In terms of the original, physical
coordinate $x_{ph}$ the new rescaling is independent of the system size:
$$\bar x = \left(\frac{2 \pi}{3}\right)^{1/2}\frac{(1-r)^{1/2}\, x_{ph}}{d}\,.$$
Equations
(\ref{E7}) for the strip state become
\begin{eqnarray}
(FZ^{\prime})^{\prime}= Q\,,  \indent Z(\bar{x}=0) = Z_0 \nonumber \\
 \mbox{and}  \indent Z^{\prime}(\bar{x}=0)=0\,, \label{strip1}
\end{eqnarray}
while the eigenvalue problem (\ref{E8}) and (\ref{E9}) reads
\begin{eqnarray}
\phi^{\prime\prime}-\left(\frac{Q_Z}{F} + \bar{k}^2\right)\phi = 0\,, \indent \phi^{\prime}
(\bar{x}=0)=0 \nonumber \\
\mbox{and} \indent \phi(\bar{x}=\eta^{1/2})=0\,. \label{E16}
\end{eqnarray}
Now the primes denote the derivatives with respect to $\bar{x}$, while $\bar{k}=k\eta^{-1/2}$ is
the new scaled wave number. Like the coordinate $\bar{x}$, the new scaled wave number $\bar{k}$ is
independent of $L_x$:
$$
\bar{k} = \left(\frac{3}{2 \pi}\right)^{1/2}\frac{k_{ph}\,d}{(1-r)^{1/2}}\,,
$$
where $k_{ph}$ is the physical wave number. The problem (\ref{strip1}) and (\ref{E16}) is
determined by two parameters: $Z_0$ and $\eta$. However, $\eta$ enters the rescaled equations only
in one place: in the last boundary condition in Eq. (\ref{E16}) where it determines the scaled
system size. If the wave function $\phi (\bar{x})$ is strongly localized in the potential well
$U(\bar{x})$ (correspondingly, the negative energy level is sufficiently deep), the results for
$\bar{k}$ and $\phi(\bar{x})$ become independent of $\eta$ at sufficiently large $\eta$. Indeed, in
this case one can safely move the boundary $\bar{x}=\eta^{1/2}$ to infinity. It is important that,
in this case, the exact form of the boundary condition at the driving wall becomes insignificant,
leading only to exponentially small corrections \cite{boundary}. This universal ``localization
regime" was discovered in Ref. \cite{LMS} that employed a different set of constitutive relations.
The criterion for localization can be obtained from the requirement that the localization length
(which is of order $\bar{k}^{-1}$) be much smaller than the (scaled) system size $\eta^{1/2}$. In
the physical units it corresponds to the short-wavelength limit of the bifurcating solution:
$k_{ph} L_x \gg 1$. To fulfill this criterion, $\bar{k}$ should be far enough from the borders of
the instability interval, where $\bar{k}$ vanishes. In the next subsection we will work in the
dilute limit and rewrite this criterion in terms of the governing parameters of the problem.

\begin{figure}[ht]
\vspace{-0.3 cm} \center{\epsfxsize=7.5 cm \epsffile{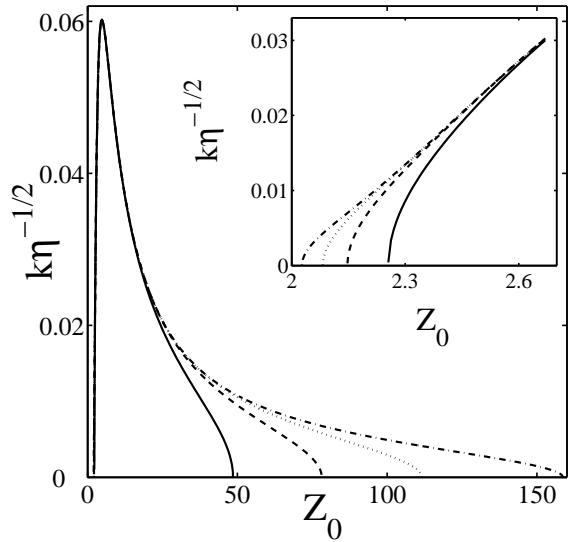}} \caption{Marginal stability curves
for different values of $\eta$, plotted in coordinates $\bar{k}=k\eta^{-1/2}$ versus $Z_0$. For a
fixed $\eta$ the strip state is stable above the respective curve and unstable below the curve.
The inset shows the splitting of the curves near the high-density stability border. The values of
$\eta$ are: $10^4$ (solid line), $2.5\cdot 10^4$ (dashed line), $5\cdot 10^4$ (dotted line) and
$10^5$ (dash-dot line). } \label{k_z}
\end{figure}

\begin{figure}[ht]
\vspace{-0.3 cm} \center{\epsfxsize=7.5 cm \epsffile{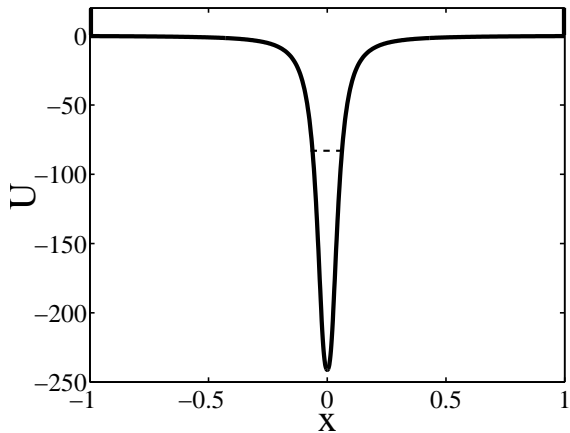}} \caption{An example of the
negative energy level $E=-k^2\simeq -83.01$ (dashed line) in the potential $U(x)$ (solid line) in
the regime of localization. The parameters are $\eta=2.5 \cdot 10^4$ and $Z_0=6$.}
\label{potential_two}
\end{figure}

\begin{figure}[ht]
\vspace{-0.3 cm} \center{\epsfxsize=7.5 cm \epsffile{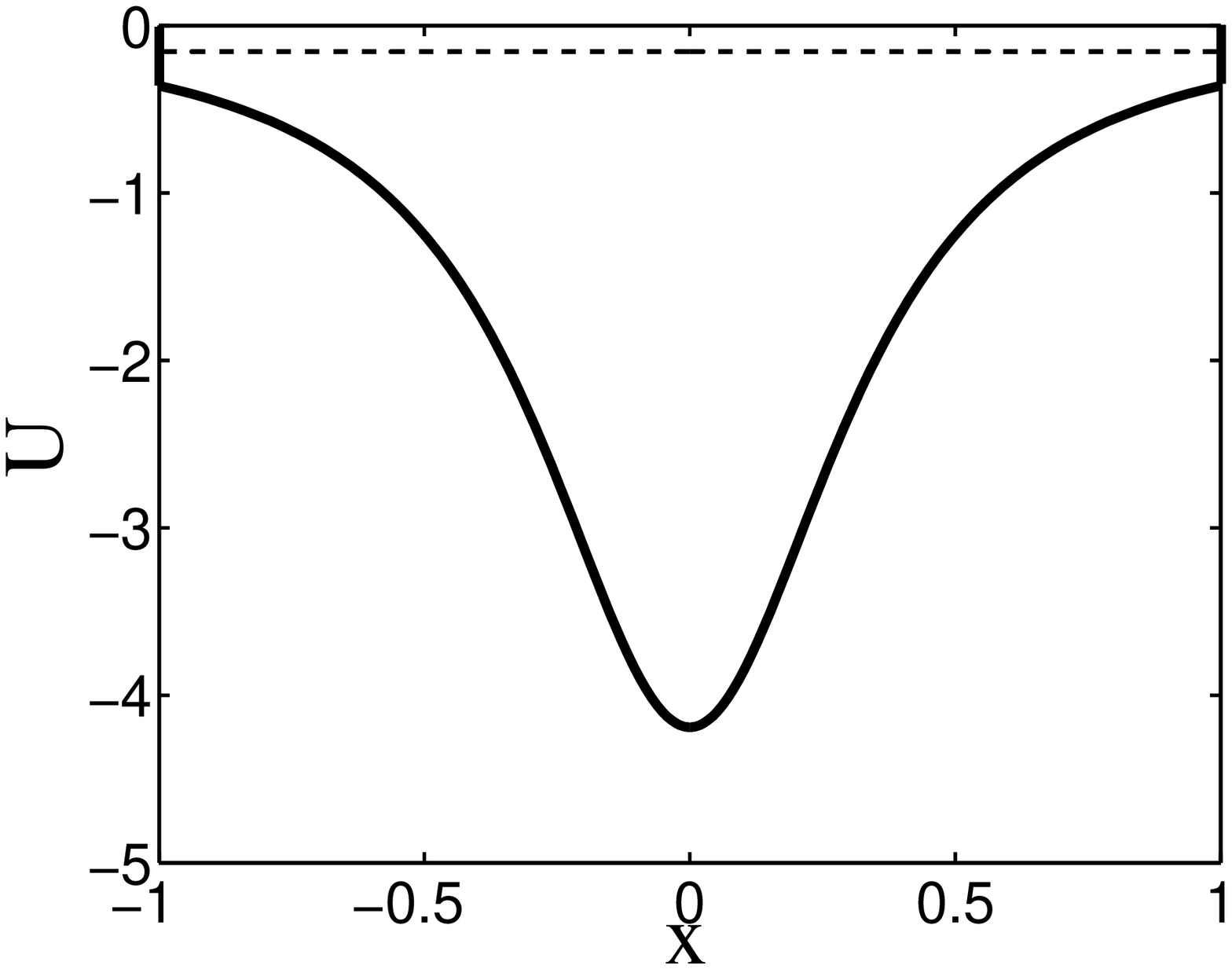}} \caption{An example of the
negative energy level $E=-k^2\simeq -0.155$ (dashed line) in the potential $U(x)$ (solid line) in
the absence of localization. The parameters $\eta=2.5 \cdot 10^4$ and $Z_0=75$ correspond to the
region close to the low-density stability border.} \label{potential_one}
\end{figure}

\begin{figure}[ht]
\vspace{-0.3 cm} \center{\epsfxsize=7.5 cm \epsffile{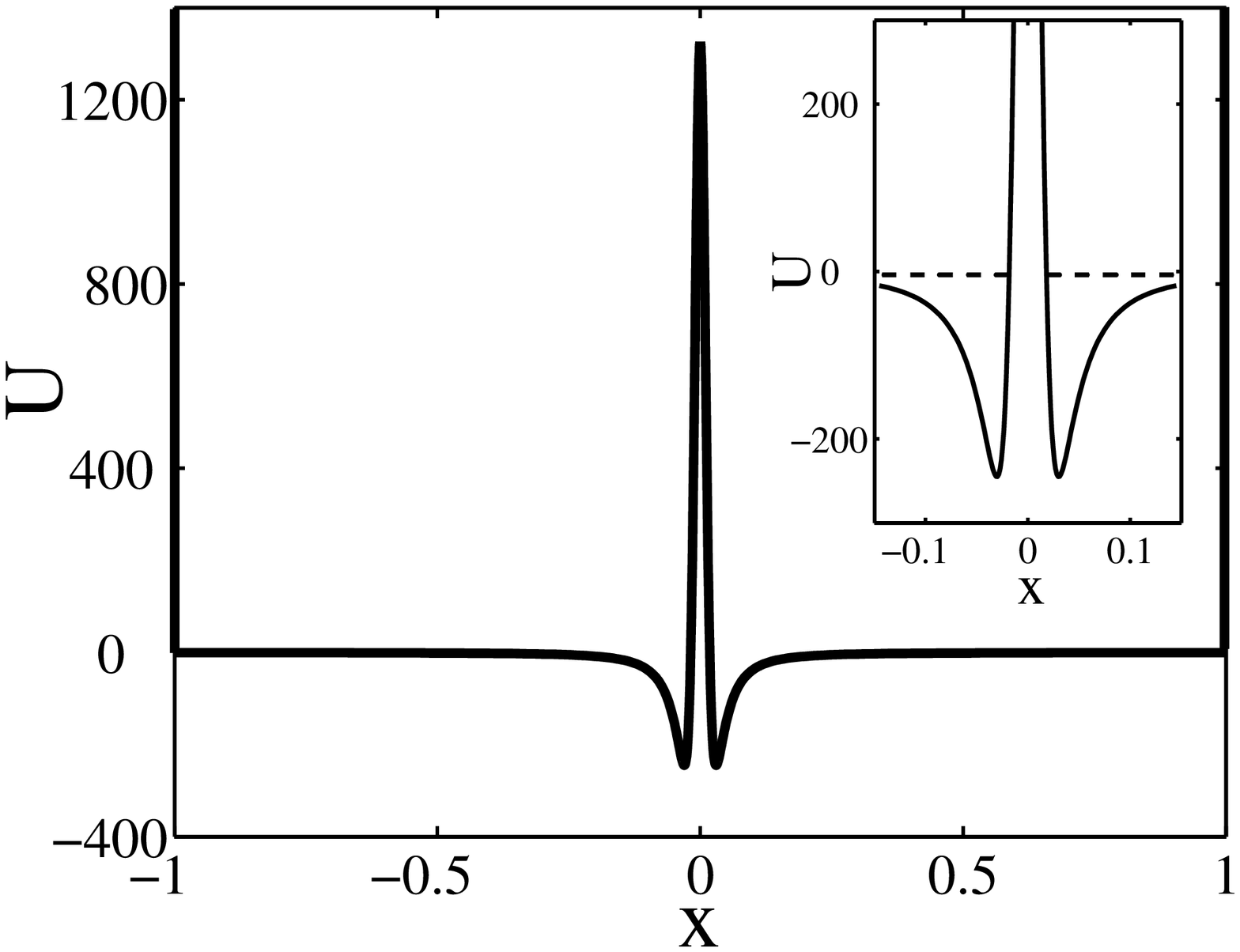}} \caption{Another example of the
negative energy level $E=-k^2\simeq -4.1$ (dashed line) in the potential $U(x)$ (solid line) in
the absence of localization. The parameters $\eta=2.5 \cdot 10^4$ and $Z_0=2.3$ correspond to the
region close to the high-density stability border.} \label{potential_three}
\end{figure}

\begin{figure}[ht]
\vspace{-0.3 cm} \center{\epsfxsize=7.5 cm \epsffile{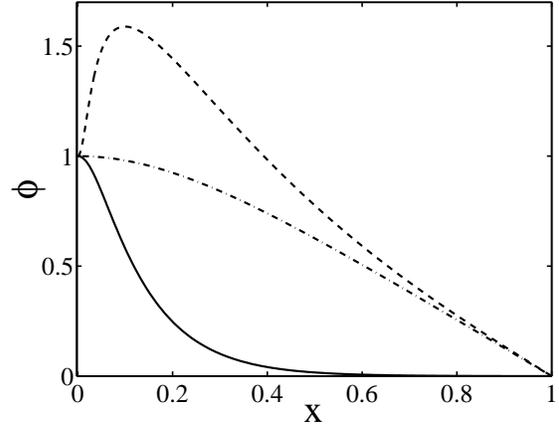}} \caption{Eigenfunctions $\phi(x)$
corresponding to the eigenvalues shown in Figs.~\ref{potential_two} (strong localization, solid
line), \ref{potential_one} (dash-dot line) and \ref{potential_three} (dashed line). The
eigenfunctions are (arbitrarily) normalized so that $\phi(0)=1$.} \label{eigenfunction}
\end{figure}

Figure~\ref{k_z} shows the marginal stability curves $\bar{k}=\bar{k}(Z_0)$ for different values of
$\eta$, obtained numerically.  Instead of the borders $f_1(\eta)$ and $f_2(\eta)$ of the
instability interval in terms of parameter $f$, the respective borders in terms of parameter $Z_0$
appear. One can see that, for large values of $\eta$, the marginal stability curves coincide in a
wide region of $Z_0$ not too close to the borders of the instability interval. This region
corresponds to strong localization. Figures~\ref{potential_two}-\ref{potential_three} show the
form of the potential (\ref{pot}) and the negative energy level $-k^2$ in three characteristic
cases (in these figures we returned to the rescaling of the coordinates and wavenumber by the
system size $L_x$). Figure~\ref{potential_two} corresponds to the region of parameters where the
energy level is deep and eigenfunction is localized. Figures~\ref{potential_one} and
\ref{potential_three} correspond to the parameter regions close to the low- and high-density
borders of the instability, respectively. There is no localization here. Notice the qualitative
change in the form of the potential near the high-density stability border.
Figure~\ref{eigenfunction} shows the respective eigenfunctions in these three cases.

\subsection{B. Marginal stability borders and physics of the instability}

\begin{figure}[ht]
\vspace{-0.3 cm} \center{\epsfxsize=7.5 cm \epsffile{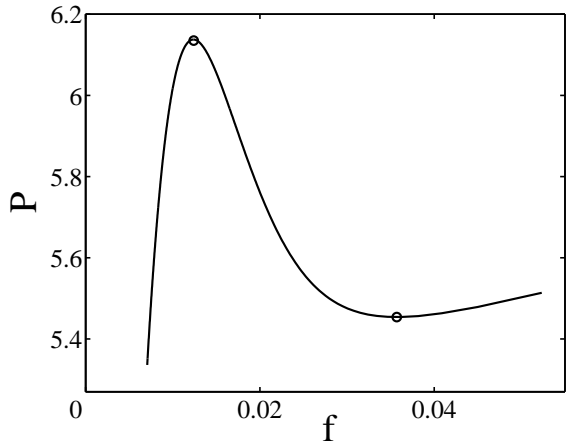}} \caption{The scaled steady-state
granular pressure $P$ versus the grain area fraction $f$ for the strip state. The two circles
correspond to the marginal stability borders $f_1$ and $f_2$. The effective lateral
compressibility of the gas is negative on the interval $(f_1, f_2)$. Parameter $\eta=10^4$.}
\label{pressure}
\end{figure}

The low- and high-density stability borders $f_1 (\eta)$ and $f_2 (\eta)$ (or respective borders in
terms of $Z_0$) are determined by the zero-eigenvalue ($\bar{k}=0$) solution of Eq.~(\ref{E16}).
This solution can be found if we know the strip solution $Z(\bar{x};Z_0)$  of Eq.~(\ref{strip1}).
Indeed, it is easy to check that function $\phi_0(\bar{x})=F\,\partial Z/\partial Z_0$ is a
solution of Eq.~(\ref{E16}) with $\bar{k}=0$, satisfying the boundary condition
$\phi^{\prime}(\bar{x}=0)$. Employing the second boundary condition $\phi(\bar{x}=\eta^{1/2})=0$
we obtain $\partial Z_1/\partial Z_0=0$, where $Z_1=Z(\bar{x}=\eta^{1/2};Z_0)$. For a given $\eta$,
this equation is an algebraic equation for $Z_0$. Our numerical results imply that this equation
has only two solutions corresponding to the low- and high-density instability borders. The
instability borders have a clear physical meaning which sheds light on the physics of the
instability. Let us consider the granular pressure $p=n T (1+2 G)$ \cite{Jenkins} of the strip
state, and introduce a scaled pressure $$ P=\frac{p}{n_c T_0} = \frac{1+ 2 G}{Z}\frac{T}{T_0}\,.
$$
As $P$ is independent of the coordinates, we can compute it at the thermal wall
$\bar{x}=\eta^{1/2}$. Here $T=T_0$ and $Z=Z_1$, so we arrive at
$$
P = \frac{1+ 2 G(Z_1)}{Z_1}=P(\eta, f)\,.
$$
Now let us compute the derivative $\partial P/\partial f$ at a constant $\eta$:
$$
\frac{\partial P}{\partial f}=\frac{\partial P}{\partial Z_1}\,\frac{\partial Z_1}{\partial
Z_0}\,\frac{\partial Z_0}{\partial f}\,.
$$
One can easily check that the first and third multipliers in the right hand side of this relation
are always negative. Therefore, the sign of $\partial P/\partial f$ is determined by the sign of
$\partial Z_1/\partial Z_0$. As we have seen, the marginal stability borders are determined by
equation $\partial Z_1/\partial Z_0=0$. Therefore,  the steady-state pressure has  its extremum
points points exactly at the points $f_1$ and $f_2$. Figure~\ref{pressure} shows an example of the
dependence $P=P(f)$ at a constant $\eta$ for the strip state, found numerically. One can see, that
$P$ is a decreasing function of $f$ within the instability interval $(f_1, f_2)$, and an increasing
function of $f$ outside the interval. The physical interpretation of these results is clear. The
presence of the anomalous (falling) part of the $P(f)$ curve indicates instability, and it is
caused by the destabilizing role of collisional heat losses \cite{Goldhirsch,plasma}. We can say
that, on the interval $(f_1, f_2)$, the granulate has \textit{negative} lateral compressibility.
At $f<f_1$ the heat losses are too small to cause instability. The presence of the high-density
border $f_2$ is caused by the finite-density corrections to the constitutive relations (that is, by
the finite size of the particles). This is in contrast to radiative condensations in gases and
plasmas, where such a stabilizing effect would be absent. Now consider a small density modulation
of the strip state with a very long wavelength: $k\rightarrow 0$. For this perturbation, the
stabilizing effect of the lateral heat conduction vanishes, and the negative compressibility
makes  the strip state unstable on the interval $(f_1, f_2)$. For a non-zero $k$, the lateral heat
conduction has a stabilizing effect. Therefore, a density modulation with too a short lateral
wavelength should be stable, as Fig.~\ref{f11} indeed shows.
\subsection{C. Dilute limit}

In this subsection we shall work in the dilute limit and derive several analytic results. We shall
see that, at large $\eta$, the dilute limit faithfully reproduces the low-density parts of the
marginal stability curves.

\subsubsection{1. Strip state and marginal stability problem}
In the dilute limit, $Z \gg 1$, Eq. (\ref{strip1}) for the strip state becomes
\begin{equation}
(Z^{3/2})^{\prime\prime} = 3 Z^{-1/2}\,, \label{E10}
\end{equation}
where the primes denote the $\bar{x}$ derivatives. The boundary conditions are
$Z^{\prime}(\bar{x}=0) = 0$ and $Z(\bar{x}=0)=Z_0$, where $Z_0$ is related to $f$ and $\eta$ by the
normalization condition $\int^{\sqrt{\eta}}_0 Z^{-1} d\bar{x} = f \eta^{1/2}$. The solution of
this problem is elementary:
\begin{equation}
\bar{x}=\frac{Z_0}{2}\left(\mbox{arccosh}\sqrt{\zeta}+\sqrt{\zeta^2-\zeta}\right)\,, \label{E11}
\end{equation}
where $\zeta=Z/Z_0$ and
\begin{equation}
Z_0= \frac{2\eta^{1/2}}{f\eta^{1/2}+\frac{1}{2}\,\sinh(2f\eta^{1/2})}\,.
\label{Z0}
\end{equation}
(Returning for a moment to the old rescaling of the coordinate, $x_{ph}/L_x \to x$, one can see
that the density profile (\ref{E11}) is determined by a \textit{single} parameter: $\xi=f
\eta^{1/2}$.) Eq. (\ref{E16}) for $\phi(\bar{x})=(\sqrt 3/2)\,Z^{1/2}(\bar{x})\,\psi_k(\bar{x})$
takes the form
\begin{equation}
\phi^{\prime\prime}-\left(\kappa-\frac{1}{\zeta^2}\right)\phi = 0,
\label{E12}
\end{equation}
where $\kappa = \bar{k}^2 Z_0^2$, and $\zeta=\zeta(\bar{x})$ is
given, in an implicit form, by Eq. (\ref{E11}). The boundary
conditions for Eq. (\ref{E12}) remain the same as in Eq.
(\ref{E16}). As $\zeta(\bar{x})$ is a monotonic function of
$\bar{x}$, we can change the independent variable in Eq.
(\ref{E12}) from $\bar{x}$ to $\zeta$. The resulting equation for
$\phi(\zeta)$ is
\begin{equation}
4(\zeta-1)\, \zeta\, \phi^{\prime\prime}+2\phi^{\prime}+(1-\kappa \,\zeta^2)\phi = 0\,, \label{EE}
\end{equation}
where the primes now denote the $\zeta$ derivatives. The function $\phi (\zeta)$ is defined on the
interval $1\le\zeta\le\zeta_1$, where $\zeta_1=Z_1/Z_0=\cosh^2 \xi$. One boundary condition is
$\phi(\zeta=\zeta_1)=0$ to which we may add an arbitrary normalization condition $\phi(\zeta=1) =
1$. An additional boundary condition, $\phi^{\prime}(\zeta=1) = (\kappa-1)/2 $, can be found from
Eq. (\ref{EE}) itself, after substituting there $\zeta=1$. This eigenvalue problem includes a
single parameter $\xi$, while $\kappa$ serves as the eigenvalue.

We have been unable to solve Eq. (\ref{EE}) analytically for a general $\kappa$. Still, several
important asymptotics can be obtained.

\subsubsection{2. Zero-energy state and stability border $f_1$}

\begin{figure}[ht]
\vspace{-0.3 cm} \center{\epsfxsize=7.5 cm \epsffile{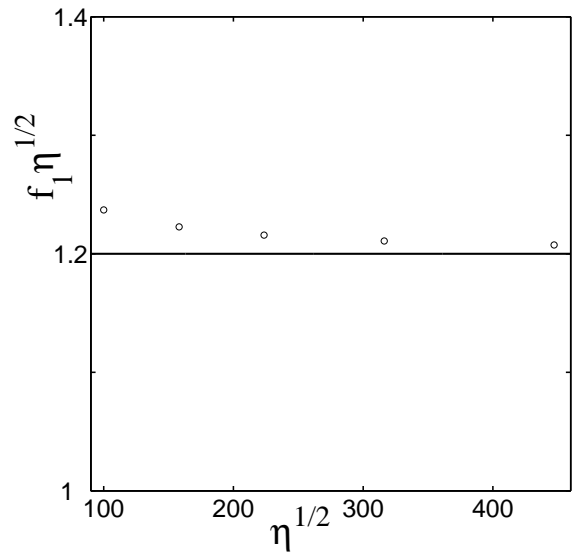}} \caption{The scaled low-density
stability boundary $\xi_1=f_1 \eta^{1/2}$ at different $\eta^{1/2}$ as found numerically from Eq.
(\ref{E8}) (circles). Solid line shows the analytical result $\xi_1=1.19968\dots$ obtained in the
dilute limit. } \label{f1}
\end{figure}

For $\bar{k}=0$ Eq. (\ref{EE}) can be solved analytically:
\begin{equation}
\phi (\zeta, \bar{k}=0)\equiv\phi_0(\zeta) =\sqrt{\zeta}-\sqrt{\zeta-1}\,\,\,
\mbox{arccosh}\sqrt{\zeta}, \label{E14}
\end{equation}
In other words, we impose a zero eigenvalue $\bar{k}=0$ and find the low-density stability border
$f_1=f_1(\eta)$ from the boundary condition $\phi_0(\zeta_1)=0$.  We obtain an algebraic equation
$\coth(\xi_1) = \xi_1$ for $\xi_1=f_1\eta^{1/2}$. Its solution is $\xi_1=1.19968\dots$. This result
explains why all marginal stability curves shown in Fig.~\ref{f11} depart (almost) from the same
point at the low-density side. Figure~\ref{f1} compares the scaled quantity $f_1 \eta^{1/2}$ at
different $\eta$, found numerically from Eq. (\ref{E8}), with this analytic prediction (a
constant). The agreement is very good for large $\eta$. As $\eta$ goes down, $f_1$ increases and
the dilute approximation starts to deteriorate.

\begin{figure}[ht]
\vspace{-0.3 cm} \center{\epsfxsize=7.5 cm \epsffile{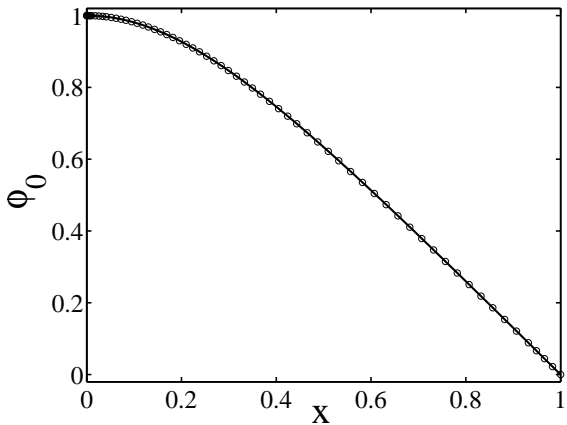}} \caption{The zero-energy
eigenfunction $\phi_0(x)$  computed numerically from Eq. (\ref{E8}) (circles) and given
analytically by Eqs.(\ref{E14}) and (\ref{E11}) (solid line). The parameters are $\eta=10^4$,
$f_{1,num}=0.0124$ (circles) and $f_{1,anal}=0.0120$ (solid line).} \label{phi}
\end{figure}

Figure~\ref{phi} compares the analytic result for the zero-energy eigenfunction $\phi_0(x)$, given
by Eqs. (\ref{E14}) and (\ref{E11}), with a numerical solution of Eq. (\ref{E8}) for $\eta=10^4$.
The coordinate $x$ in Fig.~\ref{phi} is rescaled by $L_x$. The analytic and numeric results are
obtained for slightly different values of $f_1$ (see Fig.~\ref{f1}). One can see that the agreement
is excellent.

\subsubsection{3. Short-wavelength limit: localization and universality}
\begin{figure}[ht]
\vspace{-0.3 cm} \center{\epsfxsize=7.5 cm \epsffile{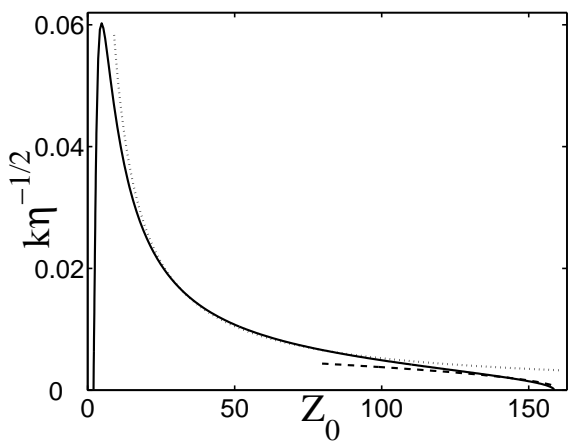}} \caption{The marginal stability
curve for $\eta=10^5$, plotted in coordinates $\bar{k}=k \eta^{-1/2}$ versus $Z_0$ (solid line)
and two dilute-limit asymptotics: the dilute-limit part $\bar{k}=A/Z_0$ of the (``universal")
short-wavelength curve  (dotted line) and the long-wavelength asymptotics (\ref{P8}) (dashed
line).} \label{ksquare_analyt}
\end{figure}

In the short-wavelength limit the system boundary $\zeta=\zeta_1=\cosh^2 \xi$ can be moved to
infinity. This requires a strong inequality $\xi=f \eta^{1/2} \gg 1$. In this limit, the
eigenvalue problem (\ref{EE}) does not include any parameter. The eigenvalue $\kappa$ should
therefore be a number of order of unity, hence $\bar{k}=A/Z_0$, with constant $A$ of order of
unity. The constant can be found numerically: $A\simeq 0.525 $. This simple result represents the
low-density limit of the ``universal" marginal stability curve, corresponding to strong
localization. Figure~\ref{ksquare_analyt} shows this asymptotics for $\eta=10^5$. One can see
excellent agreement for large enough $Z_0$, but not too close to the higher-$Z_0$ (low-density)
instability border. Near the instability border $\xi$ becomes of order of unity, and localization
breaks down.

\begin{figure}[ht]
\vspace{-0.3 cm} \center{\epsfxsize=7.5 cm \epsffile{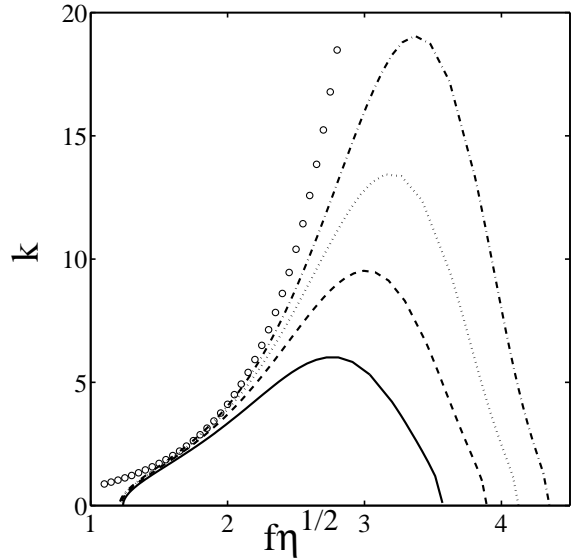}} \caption{ Marginal stability
curves for different values of $\eta$, plotted in coordinates $k$ versus $\xi = f \eta^{1/2}$. For
a fixed $\eta$, the strip state is stable above the respective curve and unstable below the curve.
The values of $\eta$ are: $10^4$ (solid line), $2.5\cdot 10^4$ (dashed line), $5\cdot 10^4$
(dotted line) and $10^5$ (dash-dot line). Also shown is the dilute-limit asymptotics
(\ref{ucurve}) of the universal marginal stability curve.} \label{kfetta}
\end{figure}

Returning to the parameters $\eta$ and $f$, and to the wave number $k=k_{ph} L_x$, we can rewrite
the asymptotics $\bar{k} = A/Z_0$ as
\begin{equation}
k=\frac{A}{2} \left[f\eta^{1/2}+\frac{1}{2}\,\sinh(2f\eta^{1/2})\right]\,. \label{ucurve}
\end{equation}
This asymptotics is shown in Fig.~\ref{kfetta}.

The asymptotics (\ref{ucurve}) is valid when $\xi=f \eta^{1/2} \gg 1$. It is easy to check that
this criterion coincides, in the dilute limit, with the localization criterion $\bar{k}^{-1} \ll
\eta^{1/2}$ discussed in Sec. 4A. On the other hand, the parameter $\xi$ should not be too large,
so that the dilute limit condition $Z_0 \gg 1$ is still satisfied, see Eq. (\ref{Z0}). These two
criteria can be rewritten as a strong double inequality for $f$: $$ \frac{1}{\eta^{1/2}} \ll f \ll
\frac{\log(8 \eta^{1/2})}{2 \eta^{1/2}}\,, $$ that can be satisfied only for extremely large
$\eta$.

\subsubsection{4. Long-wavelength limit: perturbation theory}

Close to the low-density stability border, $f-f_1 \ll f_1 = 1.19968\dots \eta^{-1/2}$ we can
assume that $\kappa \ll 1$ and solve Eq. (\ref{EE}) perturbatively. In the physical units, this
strong inequality corresponds to the long-wavelength limit: $k_{ph} L_x \ll 1$. In its turn, the
dilute limit requires $\eta \gg 1$. We substitute in Eq. (\ref{EE}) $\phi (\zeta)=\phi_0
(\zeta)+\kappa \Omega (\zeta)$, where $\phi_0$ is given by Eq.~(\ref{E14}). Neglecting the
$\kappa^2$-term in Eq. (\ref{EE}), we obtain
\begin{equation}
4 (\zeta-1)\,\zeta\Omega^{\prime\prime}+2 \Omega^{\prime}+\Omega =\zeta^2\phi_0(\zeta)\,,
\label{E17}
\end{equation}
The normalization and boundary conditions at $\zeta=1$ are $\Omega(\zeta=1)=0$ and
$\Omega^{\prime}(\zeta=1)=1/2$, respectively. The latter condition follows from Eq. (\ref{E17})
itself. Equation (\ref{E17}) can be solved analytically. With the account of the two boundary
conditions we obtain
\begin{equation}
\Omega(\zeta)=-\frac{1}{8}\,\Phi_2(\zeta)+\Phi_2 (\zeta) I_1 (\zeta)-\Phi_1 (\zeta) I_2 (\zeta),
\label{P1}
\end{equation}
where $\Phi_1(\zeta)=(\zeta-1)^{1/2}$ and $\Phi_2(\zeta)=-2\phi_0(\zeta)$. $I_1$ and $I_2$ are
indefinite integrals:
\begin{equation}
I_1=\int^{\zeta} \frac{\Phi_1 G_1}{G_2 W}\,d\zeta  \indent \mbox{and}  \indent I_2=\int^{\zeta}
\frac{\Phi_2 G_1}{G_2 W}\,d\zeta, \label{P3}
\end{equation}
where $G_1=\zeta^2\phi_0/4\,,\,G_2=\zeta(\zeta-1)$,
$W=\Phi_1\Phi^{\prime}_2-\Phi^{\prime}_1\Phi_2$ and the primes denote the $\zeta$ derivatives.
Integrals $I_1$ and $I_2$ can be evaluated in elementary functions, but the results are too
cumbersome to be presented here. The additional boundary condition [at
$\zeta=Z_1/Z_0\equiv\zeta_1$] reads $\phi(\zeta_1)=\phi_0(\zeta_1)+\kappa \Omega(\zeta_1)=0$ which
yields
\begin{equation}
\kappa (\xi)=-\phi_0(\zeta_1)/\Omega(\zeta_1)\,,  \label{P5}
\end{equation}
where $\zeta_1=\cosh^2\xi$. At the low-density stability border $f=f_1$ we have
$\xi=\xi_1=1.19968\dots .$ In the perturbative treatment, one should expand $\kappa(\xi)$ near
$\xi=\xi_1$ up to the linear term $\xi-\xi_1$. The zero-order term vanishes, and we obtain
$$\kappa(\xi)=-\frac{\Phi_2'(\xi_1)}{2\Phi_1(\xi_1)I_2(\xi_1)}\left(\xi-\xi_1\right)\,.$$
In the physical variables we have
\begin{equation}
k_{ph}L_x=\frac{\eta^{1/2}}{Z_0(\xi_1)}\left(-\frac{\Phi_2'(\xi_1)}{2\Phi_1(\xi_1)\,I_2(\xi_1)}\right)^{1/2}\left(\xi-\xi_1\right)^{1/2}.
\label{P6}
\end{equation}
As $Z_0$ is proportional to $\eta^{1/2}$, the right hand side of Eq. (\ref{P6}) is actually
independent of $\eta$. Evaluating the integral $I_2(\xi_1)$, we obtain
$I_2(\xi_1)=-0.883381\dots$. The final result is
\begin{equation}
k_{ph}L_x = 2.5115\dots\left(\xi-\xi_1\right)^{1/2}. \label{P7}
\end{equation}
Alternatively, we obtain
\begin{eqnarray}
\bar{k}(Z_0)&=&\left(\frac{\xi_1}{Z_0(\xi_1)}\right)^{3/2}\left(\frac{Z_0-Z_0(\xi_1)}{2I_2(\xi_1)}\right)^{1/2}
\nonumber \\
&=& 0.000485\dots\, (Z_0(\xi_1)-Z_0)^{1/2}\,.
\label{P8}
\end{eqnarray}
The asymptotics (\ref{P8}) is depicted in Fig. \ref{ksquare_analyt}. Close to the higher-$Z_0$
(low-density) stability border it shows good agreement with the marginal stability curve found
numerically.

\section{V. Summary and Discussion}

We determined the criteria for the spontaneous symmetry-breaking instability of the laterally
uniform granular cluster (strip state) in a prototypical driven granular gas. Working in the limit
of nearly elastic particle collisions and low or moderate densities, we employed granular
hydrodynamics with the Jenkins-Richman constitutive relations \cite{Jenkins}. The instability of
the strip state can be interpreted in terms of negative compressibility of the granulate in the
lateral direction. An important limit is found, when the marginal stability curves are independent
of the details of the boundary condition at the driving wall. In this regime the density
perturbation is exponentially localized at the elastic wall opposite to the driving wall. Working
in the dilute limit, we obtained some analytic asymptotics of the marginal stability curves.

The results of this work show that the symmetry-breaking instability predicted in Ref. \cite{LMS}
is robust and does not require very special constitutive relations. The marginal stability curves
obtained in this work are quite similar to those obtained earlier \cite{LMS} for a different set
of constitutive relations (see Fig.~\ref{delta1}). There are some quantitative differences,
however. Therefore, the instability provides a sensitive test to the accuracy of constitutive
relations.

This work was focused on the criteria of instability of the strip state. In systems sufficiently
long in the lateral direction, instability occurs in a whole range of wave numbers $k$ (below the
respective marginal stability curve). Correspondingly, multiple steady state solutions with
different $k$ are possible. In a laterally infinite system, these solutions are periodic in the
lateral coordinate. A finite system selects a finite number of wavelengths \cite{LMS}. An important
issue that was not addressed in this work is selection: what is the wavelength of the resulting
symmetry-broken cluster in an infinite, or long enough, system? The selection has dynamical
nature; this important issue is addressed elsewhere \cite{LMS2}.

Recently, the predicted symmetry-breaking instability has been observed in particle simulations
\cite{Schwager}. We hope it will be investigated in experiment, too. The experimental setting can
be of the type used by Kudrolli \textit{et al.} \cite{Kudrolli,Kudrolli2}: a system of steel
spheres, rolling on a smooth surface and driven by a rapidly vibrating side wall. The present work
(see also Ref. \cite{LMS}) provides the region of parameters where the instability can be
observed. An important issue is to eliminate the static friction between the particles and surface
that occurs far enough from the driving wall. In experiment, this is achieved by slightly inclining
the system, so that a very small gravity appears \cite{Kudrolli,Kudrolli2}. As the result, the
strip state moves down, toward the driving wall \cite{Kudrolli}. The model problem investigated in
the present work does not include gravity. We expect, however, that the symmetry-breaking
instability should persist for a non-zero gravity. In fact, a similar instability has already been
observed in particle simulations of a dilute two-dimensional granular bed fluidized by a rapidly
vibrating bottom plate \cite{Sunthar}. Under conditions of the simulations \cite{Sunthar} there
was no direct, mechanical coupling between the bottom plate vibration and collective granular
motions. Therefore, the vibrofluidized system, investigated in Ref. \cite{Sunthar}, is similar
(though not identical) to the model system driven by a thermal wall. As gravity introduces an
additional scaled parameter, the phase diagram of this type of systems should be more
complicated.  For example, it is already known that, at some values of the scaled parameters,
steady ``thermal" convection (steady state of a different type) develops both in vibrofluidized
systems \cite{Sunthar,Wildman} and in systems driven by a ``thermal" wall \cite{Ramirez,He}.
Granular hydrodynamics will be instrumental in delineating the phase diagrams of these systems in
the limit of nearly elastic collisions.

Finally, when inelasticity of the particle collisions is {\it not} small, the normal stress
difference, possible lack of scale separation and non-Gaussianity in the velocity distribution may
become important. The potential role of these effects in the symmetry-breaking instability should
be the subject of further investigations.

\section*{ACKNOWLEDGEMENTS}
We are very grateful to P.V. Sasorov and J.P. Gollub for useful discussions. This research was
supported by the Israel Science Foundation, founded by the Israel Academy of Sciences and
Humanities.


\begin{references}
\bibitem{boundary} This statement is true only for a certain class of boundary conditions:
those specifying either the heat flux, or the granular temperature at the boundaries. For
vibrofluidized systems, the boundary conditions of this type are usually obtained under two
conditions. First, the vibration amplitude of the driving plate should be much less than the mean
free path of the granulate near the driving plate. Second, the vibration frequency should be much
higher than the particle collision rate near the plate.
\end{references}
\end{document}